\documentclass[lettersize, journal]{IEEEtran}

\usepackage{algpseudocode}
\usepackage{amsfonts}
\usepackage{url}

\usepackage{subfig}
\usepackage{graphicx}

\usepackage{epsf} 
\usepackage{pdfpages}
\usepackage{booktabs}
\usepackage{array}
\usepackage[font=large]{caption}
\usepackage{verbatim}
\usepackage{todonotes}

\usepackage{amsmath}
\usepackage{amsthm}
\usepackage{tikz}
\usepackage[numbers, sort]{natbib}

\newcommand{\etal}{\textit{et al.}\xspace}
\usepackage[ruled,linesnumbered]{algorithm2e}

\usepackage{titlesec}

\begin{document}

\title{Cost Optimization for Serverless Edge Computing with Budget Constraints using Deep Reinforcement Learning}

\author{\IEEEauthorblockN{Chen Chen\IEEEauthorrefmark{1}, Peiyuan Guan\IEEEauthorrefmark{2}, Ziru Chen\IEEEauthorrefmark{3},
Amir 
Taherkordi\IEEEauthorrefmark{2}, Fen Hou\IEEEauthorrefmark{4} and Lin X. Cai\IEEEauthorrefmark{3}
  		}\\
	\IEEEauthorblockA{
        \IEEEauthorrefmark{1}Department of Computer Science and Technology, University of Cambridge, UK\\
		\IEEEauthorrefmark{2}Department of Informatics,
University of Oslo, Norway\\
        \IEEEauthorrefmark{3}Department of Electrical and Computer Engineering,
Illinois Institute of Technology, USA\\
\IEEEauthorrefmark{4}Department of Electrical and Computer Engineering, University of Macau, Macao, China\\
		Email: cc2181@cam.ac.uk;
  	peiyuang@ifi.uio.no;
        zchen71@hawk.iit.edu;
        amirhost@ifi.uio.no;
        fenhou@um.edu.mo;
        lincai@iit.edu
	}}

\maketitle

\begin{abstract}
    Serverless computing adopts a pay-as-you-go billing model where applications are executed in stateless and short-lived containers triggered by events, resulting in a reduction of monetary costs and resource utilization.
    However, existing platforms do not provide an upper bound for the billing model which makes the overall cost unpredictable, precluding many organizations from managing their budgets. 
    Due to the diverse ranges of serverless functions and the heterogeneous capacity of edge devices, it is challenging to receive near-optimal solutions for deployment cost in a polynomial time.
    In this paper, we investigated the function scheduling problem with a budget constraint for serverless computing in wireless networks.
    Users and IoT devices are sending requests to edge nodes, improving the latency perceived by users.
    We propose two online scheduling algorithms based on reinforcement learning, incorporating several important characteristics of serverless functions.
    Via extensive simulations, we justify the superiority of the proposed algorithm by comparing with an ILP solver (Midaco).
    Our results indicate that the proposed algorithms efficiently approximate the results of Midaco within a factor of 1.03 while our decision-making time is 5 orders of magnitude less than that of Midaco.
\end{abstract}

	\pagestyle{empty}  
	\thispagestyle{empty} 
\thispagestyle{empty}
\section{Introduction}\label{sec::intro}
The fast advance of artificial intelligence (AI) applications and Internet of Things (IoT) has received significant attention in recent years.
However, developers need to spend a large amount of time for the management of virtual machines and servers.
Serverless computing, also known as Function-as-a-Service (FaaS)~\cite{Yu2024}, has gained great popularity due to its elasticity and simplicity of management.
Applications can be divided into multiple functions and deployed as a serverless workflow.
The FaaS model, coupled with cloud and edge computing, is beneficial for wireless networks by deploying lightweight functions in the base stations for localized and efficient response to events, and communicating with users and IoT devices~\cite{guan2024}.
In this context, edge devices can not only collect raw data but also process data~\cite{Chen2022}, improving the response time and Quality of Service by reducing latency.

Serverless computing also raises concerns about several performance overheads.
Prior work focused some of the major ones, such as cold-start issues~\cite{Chen2023, Li2022, Yu2024}, inter-function communication~\cite{Jia2021} and state management~\cite{Xu2023a, Pfandzelter2023}.

However, one aspect has been barely investigated, namely the budget of the pay-as-you-use model adopted by serverless computing.
For many users, a significant disadvantage of a pure pay-as-you-use billing model is that the overall cost is unpredictable, which precludes many organizations from managing their budgets efficiently.
When approving their budget, organizations need to understand how much cost will be incurred by serverless services over a period of time.
This raises a legitimate concern, where cloud providers need to cap the overall price similar to the way phone companies offer monthly plans with a capped amount of usage.
Our insight is that as serverless computing is increasingly adopted by organizations, it is critical to consider users' budgets while minimizing the overall deployment cost.

To address this shortcoming, this work starts with a thorough formulation of the deployment cost with budget constraints.
In particular, we formulate a function scheduling problem, aiming to optimize the deployment cost while offering a budget option to configure in serverless computing.
Furthermore, we considered the heterogeneity and connectivity of edge networks, making the problem even more complex.
After that, we prove the proposed problem is NP-hard.
Due to the computational complexity, we carefully devised two reinforcement learning approaches based on Deep Q Networks (\textbf{DQN})~\cite{zhang2024mobile} and Proximal Policy Optimization (\textbf{PPO})~\cite{chen2022deep}, incorporating several important characteristics of serverless computing which contributes to the state of the art.

Our main contributions can be listed as follows.
\begin{itemize}
    \item We formulate a request scheduling problem in edge networks as an Integer Linear Programming (ILP) problem, jointly considering network topology, deployment cost and budget.
    We proved its NP-hardness.
    \item We carefully design two reinforcement learning-based scheduling policies, incorporating several important features of serverless computing to minimize the deployment cost while meeting fine-grained budget constraints.
    \item We conducted extensive simulations using real-word topology and traces. Experimental results justified that our algorithms approximate the results of an ILP solver Midaco~\cite{MIDACO} within a factor of 1.03 while the decision-making time is 5 orders of magnitude less than that of Midaco.
\end{itemize}

The rest of this paper is organized as follows: Section II provides a detailed review of related work, while Section III introduces the system model. In Section IV, we propose two Deep Reinforcement Learning solutions, followed by a performance evaluation in Section V. Finally, Section VI offers concluding remarks.
\section{Related Work}\label{sec::relatedwork}

Existing work investigates the function scheduling to optimize a number of different objects, including latency, deployment cost and cold starts.

Xiao \etal~\cite{Xiao2024} investigate how to reduce the system cost incurred by caching functions and selecting routes in serverless computing.
An online Lazy Caching algorithm is proposed with a worst-case competitive ratio.
Li \etal~\cite{Li2022} suggests re-purposing a warm but idle container from another function, aiming to mitigate the cold-start issue.
A container management scheme is proposed to schedule intra-function sharing without introducing new security issues.
COSE~\cite{Akhtar2020} proposes a statistical learning approach to predict the cost and execution time of serverless functions without configuration information.
Thus, COSE optimizes the configuration for running a serverless function.
These articles mainly focus on the optimization of deployment cost but overlook the budget constraint and heterogeneity of edge nodes.
Chen \etal~\cite{Chen2024} uses a probabilistic function caching algorithm inspired by Greedy-Dual-Caching, optimizing the deployment cost in serverless edge computing.

Some past work has also focused on data-intensive applications in serverless computing~\cite{Shang2023, Gu2023}.
Our work differs from this past work in that we introduce a novel budget-aware problem while fully considering the heterogeneity of edge networks and the nature of serverless computing.

Some other work~\cite{Li2023, Stojkovic2023} minimizes the startup process of serverless containers, resulting in a decrease of end-to-end latency. However, this approach reduces the startup latency but leaves the cold-start mitigation halfway.
RainbowCake~\cite{Yu2024} uses layer-wise container caching and sharing to not only reduce the startup time but also the occurrence of cold-start issues.
Unlike above works only focusing on latency, our work introduces the budget model for organizations to manage their budgets. 

There are also some work~\cite{Roy2022, Pan2022} focusing on the scheduling of serverless functions while considering the heterogeneity of edge nodes.
Nevertheless, these works do not consider the budget model and hence cannot directly apply to our problem.

To summarize, this work introduces a fine-grained budget model for serverless computing and considers the heterogeneity and connectivity of edge networks which set our work apart from existing works.
\section{System Model and Problem Formulation}\label{sec::problem}

In this paper, the deployment cost for serverless invocations consists of three components: the function switching cost, function running cost and traffic routing cost.
All symbols and variables are listed in Table~\ref{tab::var}.

\begin{table}[htbp]
	\centering
	\normalsize
	\caption{Symbols and Variables}
	\label{tab::var}
	\renewcommand\arraystretch{1}            
	\begin{tabular*}{250pt}{ll}
		\toprule	
		\textbf{Symbols and Variables} & \textbf{Description}\\
		\midrule
		$\mathcal{G} = (\mathcal{V}, \mathcal{E})$ & Physical network graph\\
            $\mathcal{V}$ & Set of edge nodes\\
            $\mathcal{E}$ & Set of links\\
            $\mathcal{N}$ & Set of function types\\
            $\mathcal{T}$ & Set of time intervals\\
            $u_n$ & The required amount of \\
            & resources for type $n$ function\\
		$U_v(t)$ & The resource capacity of node $v$\\
		$q_v^n$ & The cost of creating a type $n$ \\
            & function at node $v$\\
        $\epsilon_v^n$ & The cost of running a type $n$ \\
            & function at node $v$ for one \\
            & unit of time\\
		$d_{v,v'}$  & Cost for one unit of traffic  \\
            &between node $v$ and $v'$\\

            & function $n$ generated at node $v$\\
        & and is assigned to node $v'$\\
         $c_{n}^n$ & The cost for a request  \\ 
         & with function $n$\\
         $b_n$ & The budget for a request \\
         & with function $n$\\
         \midrule
		$x_{v}^{n}(t)$ & Binary variable to indicate \\
            & if function $n$ is assigned \\
            & to node $v$\\
		$x_{v->v'}^{n}(t)$ &   Binary variable to indicate if \\ & function $n$ offloaded to node $v'$ \\
     & from node $v$\\ 
		\bottomrule
	\end{tabular*}
\end{table}

\subsection{System model}

In this work, we consider a FaaS provider offering services in an IoT network that consists of a set of geographically distributed edge nodes, where users and IoT devices are connected to the edge nodes using wireless networks.
The set of edge nodes is denoted by $\mathcal{V}=\{1,2,...v\}$ where IoT services are executed in serverless functions.
Each edge node has a resource capacity, denoting a certain amount of hardware resources as $U_v(t)$.
Also, we use $n \in \mathcal{N}$ to represent a request that requires a type $n$ function.
Service requests are generated by end-users in the system.
We also use $t \in \mathcal{T}$ to denote the time in the system.

\subsection{Function switching cost}
First, we present the function switching cost because Launching a new function requires pulling images from a remote registry and initializing the container before serving the request.
Thus, we use $q_v^n$ to represent the switching cost of a newly created function $n$ in the edge node $v$.
\begin{equation}
    C_s = q_v^n \cdot x_v^n(t),
\end{equation}
where $x_v^n(t)$ denote a binary variable, indicating if function $n$ is assigned to node $v$.

\subsection{Function running cost}
Cost is also incurred by running a specific serverless function.
Let $\epsilon_v^n$ denote the cost of running a function $n$ in node $v$, mainly incurred by the usage of hardware resources (e.g., CPU and memory).
Then, the total function running cost is denoted by:
\begin{equation}
    C_e = \epsilon_v^n(t) \cdot x_v^n(t),
\end{equation}

\subsection{Traffic routing cost}
Also, the deployment cost is incurred by routing network traffic.
When routing traffic between two edge nodes, the cost is incurred by using the bandwidth, we use a parameter $d_{v,v'}$ to denote routing one unit of traffic from node $v$ to node $v'$.
\begin{equation}
    C_t = d_{v,v'} \cdot x_{v->v'}^n(t).
\end{equation}
where $x_{v->v'}^n(t)$ is a binary variable, denoting whether function $n$ is generated at node $v$ but offloaded to node $v'$.

\subsection{Problem formulation}
The deployment cost optimization problem with budget and capacity constraints can be formulated as follows.
\begin{equation}\label{eq:obj}
   \min \sum_{t \in \mathcal{T}} \sum_{v \in \mathcal{V}} \sum_{n \in \mathcal{N}} (C_s + C_e + C_t),
\end{equation}
s.t.
Each incoming function invocation $n$ must only be assigned to one edge node.
\begin{equation}
    \sum_{v \in \mathcal{V}} x_v^n(t) = 1, \forall n \in \mathcal{N}, \forall t \in \mathcal{T},
\end{equation}
The variable $x_v^n$ and $x_{v->v'}^n(t)$ must be binary.
\begin{equation}
   x_v^n(t), x_{v->v'}^n(t) \in [0,1], \forall v, v' \in \mathcal{V},\forall n \in \mathcal{N}, \forall t \in \mathcal{T},
\end{equation}
The total amount of required hardware resources must not exceed the capacity of each edge node.
\begin{equation}
   \sum_{n \in \mathcal{N}} u_n \cdot x_v^n(t) \leq U_v(t), \forall v \in \mathcal{V},
\end{equation}

We consider a constrained budget, denoted by $b_n \in B$, and require that the cost of each service $n$ does not exceed $b_n$.
We use $c_n$ to denote the total cost of request $n$ and hence each request should satisfy the following budget constraint:
\begin{equation}
    c_n \leq b_n, \forall n \in \mathcal{N}, \forall t \in \mathcal{T}.
\end{equation}

\subsection{NP-hardness}
We should that the Generalized Assignment Problem (GAP), which is NP-hard, can be reduced to our problem.
The GAP problem refers to allocating a number of $k \in \mathcal{K}$ tasks to a number of $J \in \mathcal{J}$ agents to minimize the overall cost.
Let $d_k$ represent the size of the task $k$ so we can map $d_k$ to the size of the function $u_n$.
Similarly, we can map an agent $j$ to an edge node $v$ so the capacity of an agent $P_j$ can be mapped to the capacity of an edge node $U_v$.
Now, if we map the cost of GAP problem to the deployment cost of the proposed problem, our problem becomes finding a scheduling solution to optimize the cost.
Thus, the GAP problem is a special case of our problem, and our problem is NP-hard.

\section{Proposed Deep Reinforcement Learning Solution}\label{sec::solution}
In this section, we present two efficient algorithms based on deep reinforcement learning (DRL).

\subsection{Markov Decision Process Formulation}
Model-free reinforcement learning (RL) is a method within dynamic programming where the probabilities of state transitions are unknown, allowing us to address Markov decision process (MDP) challenges through direct interaction and learning from the environment. However, discovering an effective control policy to solve the MDP is still a complex task. In this paper, we tackle the proposed problem using DRL. Here, the edge network is deemed as the environment, while the central controller is the learning agent. The problem is structured as an MDP, where the agent aims to learn an optimal policy, denoted as $\pi$, which maps states to actions. This setup enables the agent to make a sequence of decisions, $a^{(t)}$, based on the observed state at the current time slot, $s^{(t)}$, influencing the state in the subsequent time slot, $s^{(t+1)}$, and the reward $r{(t)}$. This interaction process can be mathematically represented as $\{s^{(t)}, a^{(t)},r^{(t)}, s^{(t+1)}, a^{(t+1)}, r^{(t + 1)}s^{(t+2)}, \ldots\}$. Consequently, we define the state $s$,  the action $a$, and the reward $r$ according to the conditions of the proposed environment as follows:

\subsubsection{State Space}
The state space represents the environment, defined as a set of states, denoted by $\mathcal{S}$. Specifically, it includes the remaining hardware capacity of each edge node, the functions hosted on each node and the generated requests at time $t$. We define the set of remaining hardware capacities across all nodes as $\Phi^{(t)}$, where $ \Phi^{(t)}= \{\varphi_v^{(t)} \mid v \in \mathcal{V}\}$, with $\varphi_v^{(t)}$ representing the hardware capacities left on node $v$. Additionally, the total number of type $n$ functions existing on node $v$ is represented by $f_v^{(n,t)}$, and the set of all functions across all nodes is denoted as $\mathcal{F}$, where $\mathcal{F}^{(t)} = \{f_v^{(n,t)} \mid v \in \mathcal{V}, n \in \mathcal{N}\}$. The generated request is represented as $\mathcal{R}^{(t)} = \{v_g^{(t)}, n^{(t)}\}$, where $v_g^{(t)}$ indicates the index of the node that generated the request, and $n^{(t)}$ specifies the type of the request generated. Therefore, the set of states $S$ at time slot $t$ can be expressed as $$S^{(t)} = \{\Phi^{(t)}, \mathcal{F}^{(t)}, \mathcal{R}^{(t)}\}.$$


\subsubsection{Action Space}
The action defines the behavior of the agent. When the agent takes an action at time $t$ following a policy $\pi$, the state transitions from the current state to a new one. We denote the action space as $\mathcal{A}$, representing the set of all possible actions. In this environment, the agent selects strategies to handle the generated request, specifically:
The node assigned to handle the request at time $t$, denoted as $v_r^{(t)}$.
Whether a new container should be generated on that node to handle the request.
Thus, we have:
\begin{align} \mathcal{A}^{(t)} = \{v_r^{(t)}, g^{(t)}\}, \end{align}
where $g^{(t)} \in \{0,1\}$ is a binary variable indicating whether a new container has been generated.

\subsubsection{Reward}
In DRL, the reward serves as a score that evaluates the performance of the action $a^{(t)}$ taken in the current state $s^{(t)}$, as guided by the policy $\pi$. The agent's objective is to maximize the expected discounted reward function. This function is defined as:
\begin{align} R_t = \mathbb{E}\left[\sum^{\infty}_{j=0} \gamma^j r(t+j)\right], \end{align}
where $\gamma \in (0, 1]$ is the discount factor. A value of $\gamma$ close to 1 indicates a preference for long-term returns, while a value close to 0 emphasizes the importance of immediate returns. Since the goal of the learning process is to minimize the budget, as specified in~\eqref{eq:obj}, we define the reward function as follows:
\begin{align}
    r{(t)} = -\left(C_s{(t)} + C_{e}{(t)} + C_t{(t)}\right).
\end{align}

\subsection{Algorithm Description}

Given that the designed action space is multi-dimensional and discrete, we employ Deep Q-Learning (DQN) and the Proximal Policy Optimization (PPO) algorithm.
We name the DQN based approach as \textbf{DFaaS} and the PPO based approach as \textbf{PFaaS}, respectively.
PPO is widely recognized as an efficient and adaptable algorithm, particularly suitable for handling high-dimensional tasks and extended time sequences, making it well-suited for our application compared to other DRL algorithms.

In this framework, the agent’s goal is to explore the action space to identify optimal strategies for handling requests. The DQN algorithm allows the agent to estimate Q-values, which represent the expected reward for each action-state pair. Through experience replay, the agent progressively learns and improves its decision-making ability, especially in high-dimensional and discrete environments.

PPO, on the other hand, employs a policy gradient approach that allows for stable updates by ensuring the policy does not diverge too far from previous steps. This is accomplished using a clipping function that limits the updates within a defined range, balancing exploration and stability. PPO’s versatility and computational efficiency make it a strong choice for managing the sequential and complex decisions required in this environment.

\subsubsection{Algorithm Workflow}

The algorithm workflow consists of several steps:

State Observation: At each time step, the agent observes the current state $s^{(t)}$, including the remaining buffer sizes and container statuses across nodes.

Action Selection: Based on the observed state and following the policy $\pi$, the agent selects an action $a^{(t)}$ from the action space $\mathcal{A}$. This action determines the node assignment $v_r$ and whether a new container should be generated.

Reward Evaluation: After executing the chosen action, the agent receives a reward $r{(t)}$, assessing the efficiency of the action in terms of resource utilization and task completion.

Policy Update: Depending on the algorithm in use:

For DQN, the Q-value table is updated using the Bellman equation, with experience replay to improve learning stability.
For PPO, policy gradients are computed, and the policy is updated by adjusting gradients, constrained within a clipping range to prevent excessive divergence.

Repeat: This cycle repeats over numerous episodes, with the agent refining its policy to maximize cumulative rewards.

\subsubsection{Algorithm Objective and Convergence}

The ultimate goal is to minimize the cost while maximizing the agent’s performance in handling requests. The algorithm aims to converge to an optimal policy $\pi^*$ that can consistently yield high rewards across varying environmental states and action sequences, effectively balancing immediate and future returns.
After obtaining the optimal policy $\pi^*$ through offline training, this policy will be deployed in the edge network to manage newly generated requests in real time.
\section{Experimental Evaluation}\label{sec::experiment}

We evaluate the performance over extensive simulation with real-world traces and topology.
We conducted the simulations on a server with 105 GB RAM and an Intel(R) Xeon(R) E5645 processor with 24 cores.

\subsection{Experiment setup}

\begin{figure}[tbhp]
    \centering
    \subfloat{\includegraphics[width=0.4\textwidth]{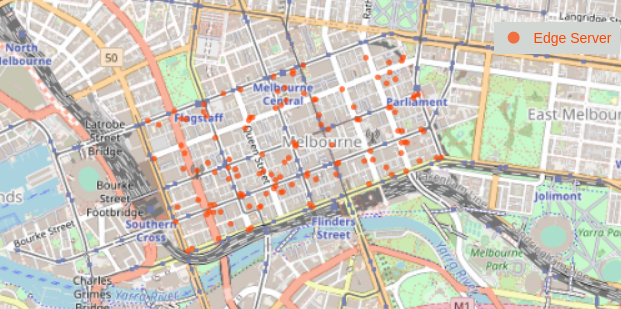}}
    \caption{Edge servers in Melbourne CBD area}
    \label{fig::topo}
\end{figure}

\textbf{Topology:} We use the EUA dataset~\cite{Lai2018} that includes 125 edge nodes from the Melbourne CBD area as shown in Figure~\ref{fig::topo}.

\textbf{Requests:} We use the Huawei serverless dataset \cite{Joosen2023} to generate serverless requests.
This dataset includes invocations of Huawei clouds for 141 days and 200 functions.

\textbf{Functions:} We selected 4 types of containers obtained from the Huawei dataset.
The memory resources allocated to each function are in [20, 250] MB.

\textbf{Edge nodes:}
To consider the heterogeneity of edge nodes,
we assume the system has 5 types of edge nodes, with CPU frequencies ranging from 2.4 to 3.6 GHz.
Also, the memory capacity is in [16, 24] GB.

\textbf{Cost parameters:}
According to \cite{Pan2022}, we assume function switching cost $q_v^n$ is inversely proportional to the CPU frequency of node $v$.
The function running cost $\epsilon_v^n$ is set to be proportional to the CPU frequency of the node.
Also, $d_v^n$ and $\epsilon_v^n$ are proportional to the function size $u_n$.

\textbf{Performance benchmarks:}
\begin{enumerate}
    \item The proposed \textbf{DFaaS} uses Deep Q-Networks to approximate the expected reward for each action in each state.

    \item The proposed \textbf{PFaaS} use Proximal Policy Optimization which optimize the agent's behavior by finding the action distribution that optimize the expected cumulative rewards.
    \item We implemented the problem in an ILP solver Midaco~\cite{MIDACO} which stochastically approximates the optimum to a mathematical problem.
MIDACO has been widely used by European Space Agency and Airbus Group and it has been proven to approach the optimum in a fast manner~\cite{MartinSchlueter2012}.
Other ILP solvers exist such as CPLEX and Gurobi but they cannot find an optimum in a polynomial time given the scale of our problem.
In Midaco, we set the maximum rounds of iteration at 200k and 300k because the results are not improved when we further increase the rounds .
\end{enumerate}

\paragraph{Box plots of deployment cost}
\begin{figure}[htbp]
    \centering
    \includegraphics[width=0.4\textwidth]{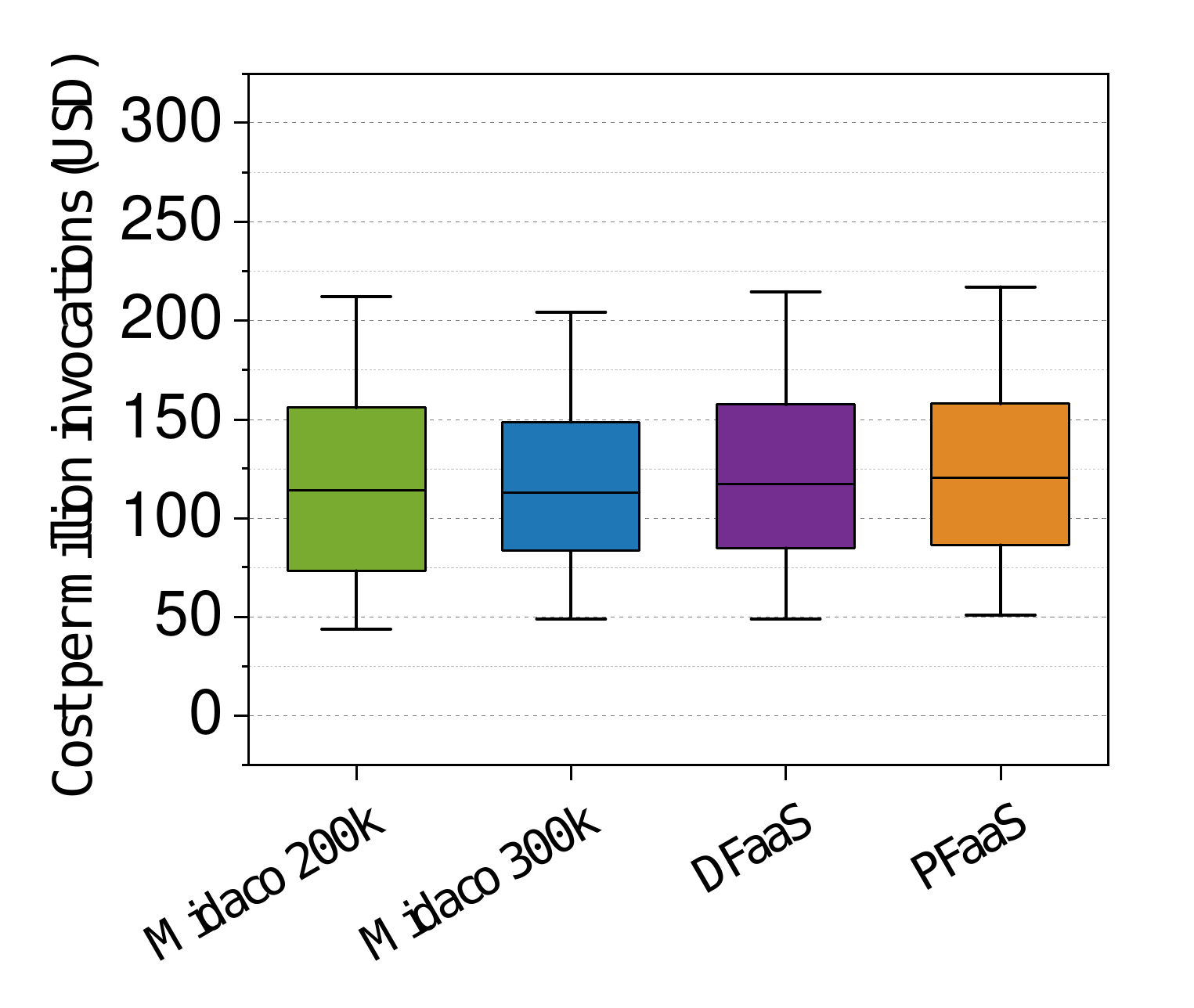}
    \caption{Box plots of deployment cost}
    \label{fig::box}
     \vspace{-3mm}
\end{figure}

In Figure~\ref{fig::box}, we present the box plots of Midaco with 200k rounds and 300k rounds, DFaaS and PFaaS for deployment costs in US dollars (USD).
The middle lines represent median values, the bottom and top of each ``box'' are 25th and 75th percentiles.
Also, the bottom and top whiskers denote the 5th and 95th percentile of the deployment costs.

We notice that Midaco with 300k rounds receives the best performance at $\$$204.26 for the 95th percentile cost where that of Midaco with 200k rounds is $\$$212.09.
This is not surprising because more iteration rounds in Midaco can lead to better results.
The proposed DFaaS and PFaas approaches yield $\$$214.39 and $\$$217.09 at 95th percentile of deployment cost.
The results shows that DFaaS and PFaaS achieves a performance within a factor of 1.05 and 1.06 to Midaco, respectively.
The results justify that DFaaS and PFaaS can efficiently approximate the results achieved by Midaco.

\paragraph{Average cost and acceptance rate}
\begin{figure}[htbp]
    \centering
    \includegraphics[width=0.4\textwidth]{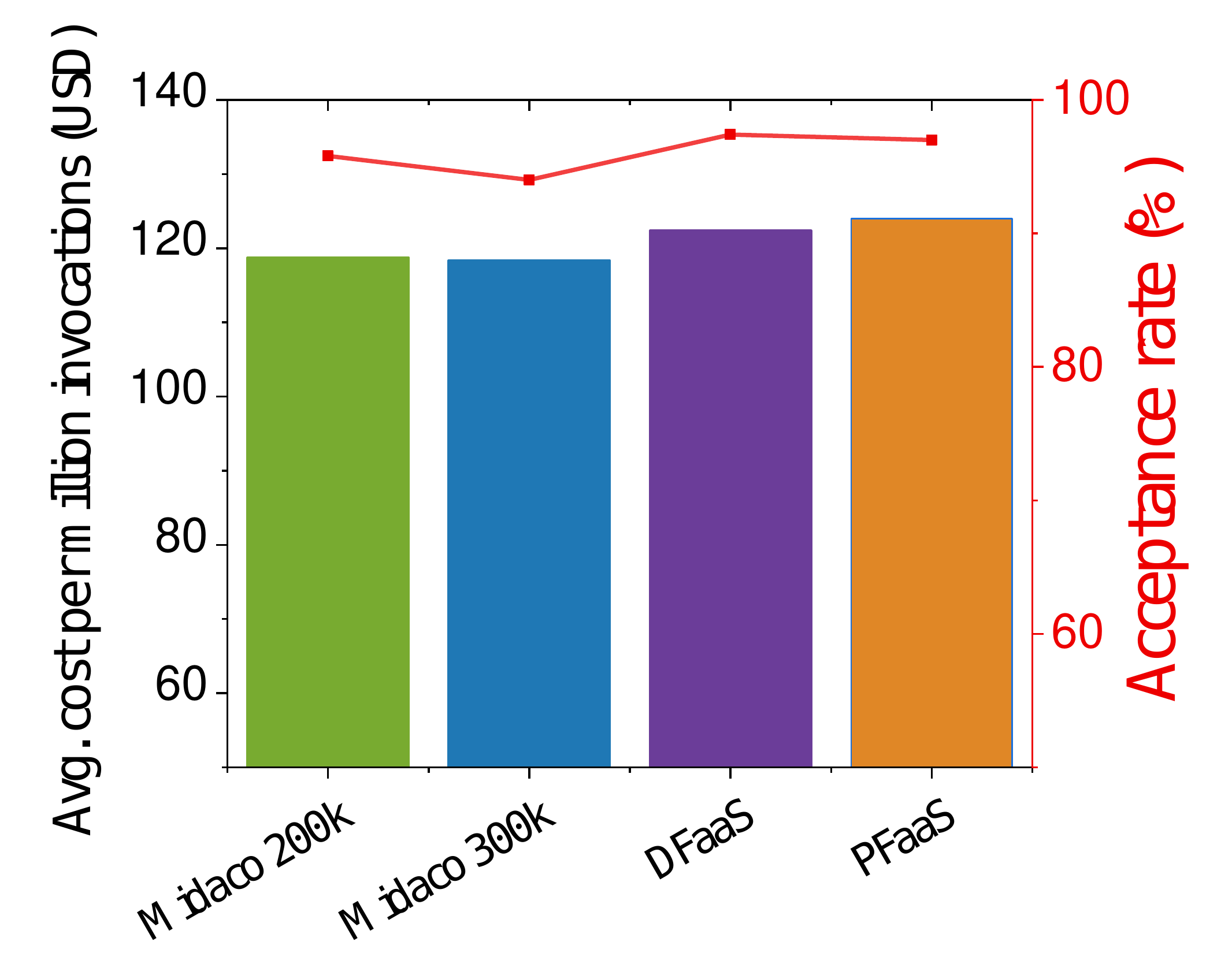}
    \caption{Average cost and acceptance rate}
    \label{fig::avg}
     \vspace{-3mm}
\end{figure}

Figure~\ref{fig::avg} illustrates the average deployment costs and the acceptance rate of the benchmarks.
The acceptance rate is the ratio between the number of deployed requests and the number of total requests.

We observe that the average deployment costs of DFaaS and PFaaS are $\$$122.35 and $\$$124.00, respectively.
In contrast, the average deployment cost of Midaco at 200k and 300k rounds are $\$$118.73 and $\$$118.35.
The results first justify that Midaco shows negligible improvement when we increase the maximum rounds from 200k times to 300k times.
Second, DFaaS and PFaaS can efficiently approximate the optimum received by Midaco within a factor of 1.03.

Figure~\ref{fig::avg} also presents the acceptance rate of different approaches.
It is not surprising that DFaaS and PFaaS achieves better performance than Midaco.
The rational is that learning-based approach can more efficiently explore the solution space and hence manages to accommodate more requests.

\paragraph{Decision-making time}

\begin{figure}[htbp]
    \centering
    \includegraphics[width=0.4\textwidth]{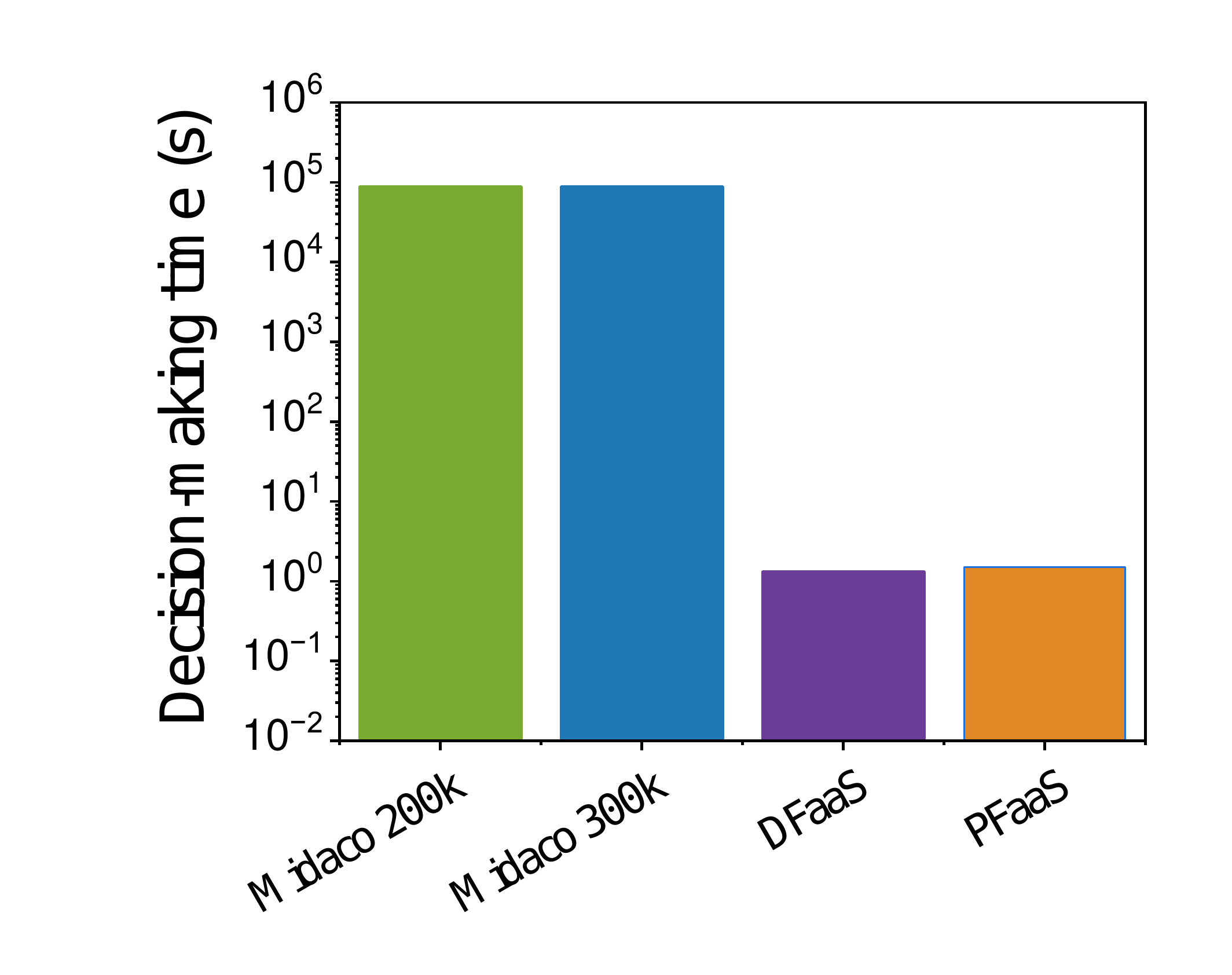}
    \caption{Decision-making time}
    \label{fig::TIME}
     \vspace{-3mm}
\end{figure}

Figure~\ref{fig::TIME} shows  the decision-making time of different approaches.
The decision-making time is the cumulative time of making decisions for all requests. For example, in DFaaS and PFaaS, the decision-making time is the time to inference the solutions.

As expected, DFaaS and PFaaS only uses 1.3 and 1.5 seconds while Midaco uses over 24 hours.
In other words, the decision-making time of the proposed approaches is 5 orders of magnitude less than Midaco.
The results justify that the proposed approaches are suitable for online decision-making although Midaco receives better performance in deployment cost with a trade-off in decision-making time.
\section{Conclusion}\label{sec::conclusion}

In this paper, we have studied the function scheduling problem for wireless networks with a budget constraint, facilitating organizations to estimate their costs under a budget.
We considered the function switching cost, running cost and routing cost and prove the problem is NP-hard.
Due to the complexity of computation, we proposed two approaches based on DQN and PPO to make online decisions.
We compared the proposed approaches with Midaco which is an ILP-solver to efficiently approximate the optimal solutions.
Extensive simulation results justify that the proposed approach receives superior performance within a factor of 1.03 compared to Midaco.
The decision-making time of the proposed approach is 5 orders of magnitude less than that of Midaco, indicating the proposed approaches are suitable for online decision-making.

Future works include implementing DFaaS and PFaaS in a commercial serverless platform to justify the efficacy.
It is also promising to apply the proposed approaches to examine other performance metrics such as latency.
This can be done by modifying the reward function in the proposed approaches.

\section*{Acknowledgment}
This work
has been partially supported by the Norwegian Research Council under
Grant 262854/F20 (DILUTE project) and 
Grant 322473 (AirQMan project);
and the University of Macau (Project No. MYRG-GRG2023-00200-FST and Conference Grant with Reference No. CG-FST-2025)

\bibliographystyle{unsrt}
{\footnotesize
\bibliography{reference}}

\begin{thebibliography}{10}

\bibitem{Yu2024}
Hanfei Yu, Rohan Basu~Roy, Christian Fontenot, Devesh Tiwari, Jian Li, Hong
  Zhang, Hao Wang, and Seung-Jong Park.
\newblock Rainbowcake: Mitigating cold-starts in serverless with layer-wise
  container caching and sharing.
\newblock In {\em Proceedings of the 29th ACM International Conference on
  Architectural Support for Programming Languages and Operating Systems, Volume
  1}, ASPLOS '24, page 335–350, New York, NY, USA, 2024. Association for
  Computing Machinery.

\bibitem{guan2024}
Peiyuan Guan, Chen Chen, Ziru Chen, Lin~X Cai, Xing Hao, and Amir Taherkordi.
\newblock Context-aware container orchestration in serverless edge computing.
\newblock {\em arXiv preprint arXiv:2408.07536}, 2024.

\bibitem{Chen2022}
Chen Chen, Lars Nagel, Lin Cui, and Fung~Po Tso.
\newblock B-scale: Bottleneck-aware vnf scaling and flow routing in edge
  clouds.
\newblock In {\em 2022 IEEE Symposium on Computers and Communications (ISCC)},
  pages 1--6, 2022.

\bibitem{Chen2023}
Chen Chen, Lars Nagel, Lin Cui, and Fung~Po Tso.
\newblock S-cache: Function caching for serverless edge computing.
\newblock In {\em Proceedings of the 6th International Workshop on Edge
  Systems, Analytics and Networking}, EdgeSys '23, page 1–6, New York, NY,
  USA, 2023. Association for Computing Machinery.

\bibitem{Li2022}
Zijun Li, Linsong Guo, Quan Chen, Jiagan Cheng, Chuhao Xu, Deze Zeng, Zhuo
  Song, Tao Ma, Yong Yang, Chao Li, and Minyi Guo.
\newblock Help rather than recycle: Alleviating cold startup in serverless
  computing through {Inter-Function} container sharing.
\newblock In {\em 2022 USENIX Annual Technical Conference (USENIX ATC 22)},
  pages 69--84, Carlsbad, CA, July 2022. USENIX Association.

\bibitem{Jia2021}
Zhipeng Jia and Emmett Witchel.
\newblock Nightcore: efficient and scalable serverless computing for
  latency-sensitive, interactive microservices.
\newblock In {\em Proceedings of the 26th ACM International Conference on
  Architectural Support for Programming Languages and Operating Systems},
  ASPLOS '21, page 152–166, New York, NY, USA, 2021. Association for
  Computing Machinery.

\bibitem{Xu2023a}
Zichuan Xu, Lizhen Zhou, Weifa Liang, Qiufen Xia, Wenzheng Xu, Wenhao Ren,
  Haozhe Ren, and Pan Zhou.
\newblock Stateful serverless application placement in mec with function and
  state dependencies.
\newblock {\em IEEE Transactions on Computers}, 72(9):2701--2716, 2023.

\bibitem{Pfandzelter2023}
Tobias Pfandzelter and David Bermbach.
\newblock Enoki: Stateful distributed faas from edge to cloud.
\newblock In {\em Proceedings of the 2nd International Workshop on Middleware
  for the Edge}, MiddleWEdge '23, page 19–24, New York, NY, USA, 2023.
  Association for Computing Machinery.

\bibitem{zhang2024mobile}
Chaoyue Zhang, Bin Lin, Ziru Chen, Lin~X Cai, and Jianli Duan.
\newblock Mobile edge deployment and resource management for maritime wireless
  networks.
\newblock {\em IEEE Transactions on Vehicular Technology}, 2024.

\bibitem{chen2022deep}
Ziru Chen, Ran Zhang, Lin~X Cai, Yu~Cheng, and Yong Liu.
\newblock A deep reinforcement learning based approach for noma-based random
  access network with truncated channel inversion power control.
\newblock In {\em ICC 2022-IEEE International Conference on Communications},
  pages 1835--1840. IEEE, 2022.

\bibitem{MIDACO}
Midaco-solver.
\newblock \url{https://www.midaco-solver.com/}, 2024.

\bibitem{Xiao2024}
Ke~Xiao, Song Yang, Fan Li, Liehuang Zhu, Xu~Chen, and Xiaoming Fu.
\newblock Making serverless not so cold in edge clouds: A cost-effective online
  approach.
\newblock {\em IEEE Transactions on Mobile Computing}, 23(9):8789--8802, 2024.

\bibitem{Akhtar2020}
Nabeel Akhtar, Ali Raza, Vatche Ishakian, and Ibrahim Matta.
\newblock Cose: Configuring serverless functions using statistical learning.
\newblock In {\em IEEE INFOCOM 2020 - IEEE Conference on Computer
  Communications}, pages 129--138, 2020.

\bibitem{Chen2024}
Chen Chen, Manuel Herrera, Ge~Zheng, Liqiao Xia, Zhengyang Ling, and Jiangtao
  Wang.
\newblock Cross-edge orchestration of serverless functions with probabilistic
  caching.
\newblock {\em IEEE Transactions on Services Computing}, 17(5):2139--2150,
  2024.

\bibitem{Shang2023}
Xiaojun Shang, Yingling Mao, Yu~Liu, Yaodong Huang, Zhenhua Liu, and Yuanyuan
  Yang.
\newblock Online container scheduling for data-intensive applications in
  serverless edge computing.
\newblock In {\em IEEE INFOCOM 2023 - IEEE Conference on Computer
  Communications}, pages 1--10, 2023.

\bibitem{Gu2023}
Rong Gu, Xiaofei Chen, Haipeng Dai, Shulin Wang, Zhaokang Wang, Yaofeng Tu,
  Yihua Huang, and Guihai Chen.
\newblock Time and cost-efficient cloud data transmission based on serverless
  computing compression.
\newblock In {\em IEEE INFOCOM 2023 - IEEE Conference on Computer
  Communications}, pages 1--10, 2023.

\bibitem{Li2023}
Yuepeng Li, Deze Zeng, Lin Gu, Mingwei Ou, and Quan Chen.
\newblock On efficient zygote container planning toward fast function startup
  in serverless edge cloud.
\newblock In {\em IEEE INFOCOM 2023 - IEEE Conference on Computer
  Communications}, pages 1--9, 2023.

\bibitem{Stojkovic2023}
Jovan Stojkovic, Tianyin Xu, Hubertus Franke, and Josep Torrellas.
\newblock Specfaas: Accelerating serverless applications with speculative
  function execution.
\newblock In {\em 2023 IEEE International Symposium on High-Performance
  Computer Architecture (HPCA)}, pages 814--827, 2023.

\bibitem{Roy2022}
Rohan~Basu Roy, Tirthak Patel, and Devesh Tiwari.
\newblock Icebreaker: warming serverless functions better with heterogeneity.
\newblock In {\em Proceedings of the 27th ACM International Conference on
  Architectural Support for Programming Languages and Operating Systems},
  ASPLOS '22, page 753–767, New York, NY, USA, 2022. Association for
  Computing Machinery.

\bibitem{Pan2022}
Li~Pan, Lin Wang, Shutong Chen, and Fangming Liu.
\newblock Retention-aware container caching for serverless edge computing.
\newblock In {\em IEEE INFOCOM 2022 - IEEE Conference on Computer
  Communications}, pages 1069--1078, 2022.

\bibitem{Lai2018}
Phu Lai, Qiang He, Mohamed Abdelrazek, Feifei Chen, John Hosking, John Grundy,
  and Yun Yang.
\newblock Optimal edge user allocation in edge computing with variable sized
  vector bin packing.
\newblock In Claus Pahl, Maja Vukovic, Jianwei Yin, and Qi~Yu, editors, {\em
  Service-Oriented Computing}, pages 230--245, Cham, 2018. Springer
  International Publishing.

\bibitem{Joosen2023}
A.~Joosen, A.~Hassan, M.~Asenov, R.~Singh, L.~Darlow, J.~Wang, and A.~Barker.
\newblock How does it function? characterizing long-term trends in production
  serverless workloads.
\newblock In {\em Proceedings of the 2023 ACM Symposium on Cloud Computing},
  SoCC '23, page 443–458, New York, NY, USA, 2023. Association for Computing
  Machinery.

\bibitem{MartinSchlueter2012}
Matthias~Gerdts Martin~Schlüter and Jan-J. Rückmann.
\newblock A numerical study of midaco on 100 minlp benchmarks.
\newblock {\em Optimization}, 61(7):873--900, 2012.

\end{thebibliography}

\end{document}